# Observation of reversible orbital angular momentum transfer based on photon-phonon coupling

Zhihan Zhu[1,2], Wei Gao[1,2*], Chunyuan Mu[2] and Hongwei Li[1,2]

*1. The higher educational key laboratory for Measuring & Control Technology and Instrumentation of Heilongjiang Province, Harbin University of Science and Technology, Harbin 150001, China*

*2. Institute of photonics and optical fiber technology, Harbin University of Science and Technology, Harbin 150001, China*

*3. Department of Physics and Max Planck Centre for Extreme and Quantum Photonics, University of Ottawa, Ottawa, Ontario K1N 6N5, Canada*

*e-mail: wei_g@163.com

**Orbital angular momentum (OAM) has gained great interest due to its most attractive feature of high dimensionality, and several ground-breaking demonstrations in communication based on OAM multiplexing have been carried out. Accordingly, a rapid data-density growth from OAM multiplexing has posed a great challenge to the signal-processing layer. Meanwhile, in another area, optical signal-processing circuit based on photon-phonon conversion has received considerable attention and made rapid progress. Here, with an aim of finding the intersection between OAM multiplexing and photon-phonon conversion, we report on the observation of reversible OAM photon-phonon conversion. A specific OAM state can be flexibly and controllably interconverted between photonic and phononic domains via Brillouin photon-phonon coupling within the decay time of acoustic signal, in which OAM and spin angular momentum are independently conserved. Our result demonstrates the controllable OAM transfer between photons and phonons, shows the potential of using OAM multiplexing to extend the capacity of photon-phonon conversion based signal-processing scheme, and may trigger the development of OAM-multiplexed photon-phonon circuit.**

## 1. INTRODUCTION

Orbital angular momentum (OAM), a photonic degree of freedom with inherent multi-dimension shown by Allen and co-workers [1-3], has made rapid progress in light manipulation, enhanced imaging, high-capacity communication and memory [4-12] due to its unique profile, ranging from light to radio waves, and even electrons and plasma [13-15]. Total photonic angular momentum can be divided into spin angular momentum (SAM) and OAM in paraxial approximation. SAM and OAM are separately conserved when a light beam propagates in vacuum or a homogeneous and isotropic medium [16, 17], providing a basis for using OAM beams to communicate information over a long distance. In comparison, information processing usually requires some interactions between signals. For OAM beams, these interactions can be well implemented by nonlinear optical processes both in classical and quantum domains, such as parametric down conversion, second harmonic generation and four-wave mixing [18-23], which are more applicable for matching signals' frequency interface [19, 23]. Besides, light-atom interactions afford some key protocols for processing optical information in quantum domain [12], but these interactions are difficult to be applied in classical realm, especially for on-chip level. In general, the signal-processing functions based on above interactions are far from the demand of OAM-multiplexing. Additionally, another nonlinear process for concern is Brillouin interaction, an inelastic scattering of light from sound, which has received increasing attention with an aim of harnessing reversible photon-phonon conversion to perform signal processing [24, 25]. Compared with other nonlinear interactions, Brillouin process affords more signal-processing functions due to involving acoustic phonons with low velocity and long-coherence time, and has a good compatibility with silicon circuit. Some related demonstrations such as light storage, signal processing and Brillouin scattering induced transparency have been carried out in silicon photon-phonon circuit or silica resonator recently [24-28].

To our knowledge, phonons carry no SAM due to its nature of electrostatic oscillations, but it is worth noting that they can carry OAM by forming acoustic vortices in the propagation medium, in other words, phonons' OAM coincides with their total angular momentum. In theory, the exchange of OAM between electromagnetic and plasmon/phonon wave has been studied in Ref. [29]. In experiment, the OAM transfer from acoustic wave to photons has been observed in fiber via acousto-optic interaction [30]. More interestingly, we found that the photonic OAM conservation was violated in stimulated Brillouin amplification (SBA) recently, and we proposed an assumption of phonon carrying OAM to explain this phenomenon [31]. This assumption indicates that OAM beam involved SBA may be a feasible method to realize the prediction of exciting specific OAM phonon from outside in Ref. [29]. And on this basis, we believe the excited OAM phonon, if it really exists, can be probed via photon-phonon conversion as well. Consequently, a new insight is naturally gained once deliberating on above assumption and the requirement of signal processing in OAM-multiplexing system, i.e., whether can we achieve a high-capacity and flexible information-processing technique by extending reversible photon-phonon conversion scheme into OAM-multiplexing area. Here, to demonstrate the physical mechanism of this proposal, we report on the observation of reversible OAM photon-phonon conversion via backward-Stimulated Brillouin scattering (SBS) process. The OAM conservation relations in SBA and Brillouin acoustic parametric amplification (BAPA) are confirmed. More specifically, a well-defined OAM state can be converted into phonon beam via SBA, and be reconverted back to optical domain via BAPA within the decay time of coherent phonons. SAM and OAM are independently conserved in these processes due to phonons only carrying OAM. Beyond the fundamental significance, this demonstration reveals a great potential use of photon-phonon conversion for information processing in OAM-multiplexing system.

## 2. THEORY BACKGROUND

Backward-SBS process, used in this proof-of-principle study, can be summarized into two quasi-parametric down-conversions, i.e., SBA and BAPA. In SBA, a pump beam and a Stokes frequency-shift seed beam counter-propagate in a medium, the seed is amplified by pump and a coherent phonon beam is simultaneously generated as an idler field. The energies and the momenta of two optical fields and phonon field in SBA must satisfy conservation relations of $\omega_p - \omega_s = \Omega$ and $k_p - k_s = q$, respectively. Here $\omega_p(k_p)$, $\omega_s(k_s)$ $\omega_s(k_s)$ and $\Omega(q)$ are the energy (momentum) of pump, seed and phonon field, respectively. In BAPA, a probe beam co-propagates with and enhances a coherent phonon beam usually generated by SBA or focused-SBS. A major difference from SBA is that here the role of phonon beam is a seed field of down-conversion, and a Stokes frequency-shift light beam is generated as an idler field. Indeed, the physical nature of so-called Brillouin dynamic grating in distributed fiber sensor domain is just a BAPA process, where the coherent phonon (dynamic grating) is generated by SBA [32]. The energy and momentum conservation required by BAPA are $\omega_b - \omega_k = \Omega$ and $k_b - k_k = q$, where $\omega_b(k_b)$, and $\omega_k(k_k)$ are the energy (momentum) of probe and Stokes field, respectively. It is noted that the momentum conservation relations $k_p - k_s = k_b - k_k = q$ include the angular momentum as well, and considering the fact that phonons carry no SAM, thus the SAM conservation for SBA and BAPA are $S_p = S_s$ and $S_k = S_b$ respectively. Here, $S_x$ is the SAM of corresponding optical fields. That is to say, the SAM conservation in photon-phonon coupling only involves optical fields. More specifically, the pump and seed should be set at same linear or opposite circular polarizations in SBA; while in BAPA, any polarized probe beam can interact with coherent phonons and generates a same linear or opposite circular polarized Stokes beam.

Next, we first discuss OAM conservation relation in SBA for deducing the phonon OAM state excited by SBA. Then, in order to intuitionally confirm the existence of SBA-excited OAM phonons, we discuss OAM conservation relation in BAPA for deducing the Stoke beam's OAM state extracted from phonons. Firstly, for obtaining the phonon OAM state excited by SBA, both consider the interaction picture Hamiltonian of SBA and the requirement of OAM conservation at *x*-axis as follows

$$H = \hbar\kappa(a_p a_s^\dagger \rho^\dagger + a_p^\dagger a_s \rho) \quad \textbf{(1)}$$

$$L_x\left|\ell_p\right\rangle = L_x(\left|\ell_s\right\rangle + \left|\ell_\rho\right\rangle) \quad \textbf{(2)}$$

where $\rho$ and $a_p$ ( $a_s$ ) are the boson annihilation operators for phonon and pump (seed) photon, respectively, and $\kappa$ is a coupling constant, $L_x$ is OAM operator for *x*-axis, $\left|\ell_\rho\right\rangle$ and $\left|\ell_p\right\rangle$ ($\left|\ell_s\right\rangle$) are OAM states of phonon and pump (seed), respectively. By removing the operator $L_x$ in Eq. (2), we obtain the SBA-excited phononic OAM state at *x*-axis is

$$\left|\ell_\rho\right\rangle = \left|\ell_p + \ell_s\right\rangle \quad \textbf{(3)}$$

which also describes the OAM transfer from photons to phonons. Notice that, here the OAM transfer is driven by the electrostriction induced by two frequency-detuned beams, and this transfer can also occur with only vortex pump presence (see additional discussion in Supplement 1).

Secondly, for obtaining the OAM conservation relation in BAPA, both consider the interaction picture Hamiltonian of BAPA and the requirement of system OAM conservation expressed as

$$H = \hbar\kappa(a_b a_k^{\dagger}\rho^{\dagger} + a_b^{\dagger}a_k\rho\ ) \quad \textbf{(4)}$$

$$L_x|\ell_b\rangle = L_x(|\ell_k\rangle + |\ell_\rho\rangle) \quad \textbf{(5)}$$

where $a_b$ ($a_k$) and $|\ell_b\rangle$ ($|\ell_k\rangle$) are the boson annihilation operators for probe (Stokes) photons and the OAM states of them respectively. Since the SBA and BAPA are connected by SBA-excited phonon field, and considering the fact that phonon state can only be indirectly observed via their induced effect on light, therefore, we are working to obtain the relation of photon OAM states in the SBA and BAPA by eliminating phonon state $|\ell_\rho\rangle$ from Eqs. (2), (3) and (5) is

$$L_x|\ell_k\rangle = L_x(|\ell_b\rangle - |\ell_p\rangle + |\ell_s\rangle) \quad \textbf{(6)}$$

The behavior of photon OAM states shown in Eq. (6) looks like a four-photon process, this is why the connected two quasi-parametric down conversions were called Brillouin enhanced four-wave mixing (BEFWM). It should be note that the name of BEFWM is not appropriate, because it is not a four-wave mixing but two independent three-wave mixing processes sharing the same phonon field of which is excited in SBA and enhanced in BAPA. By removing the operator in Eq. (6), we obtain an OAM relation between the output idler Stokes beam and the input optical beams (pump, seed and probe) given by

$$|\ell_k\rangle = |\ell_p + \ell_s - \ell_b\rangle \quad \textbf{(7)}$$

Note, the quantity of OAM transfer from phonons to photons in BAPA is opposite to the increment of phonon OAM, i.e., $-\ell_\rho$ per photon, and it is to expected that the OAM selection rules shown in Eq. (6) and (7) are equivalent version of $k_p - k_s = k_b - k_k = q$ .

**3. EXPERIMENT**

Now, we experimentally demonstrate the reversible OAM photon-phonon conversion based on above discussion. Figure 1(a) shows the schematic diagram of experimental setup, a simple method which can be easily duplicated in any optical laboratory. A P-polarized pump beam and a P-polarized Stokes-frequency-shift seed beam quasi-collinearly interact in coupling cell to excite a coherent phonon beam via SBA. Then, an S-polarized probe beam, collinearly counter-propagating with seed, interacts with the SBA-excited phonons via BAPA, and meanwhile generates an S-polarized Stokes beam which collinearly counter-propagates with pump. The Stokes beam carries the phonon's information and outputs from PBS3 (see vector diagram at right bottom of Fig. 1(a)). Usually, there are two methods to insulate the probe from interacting with seed, one of them is called frequency-decoupling, a frequency-difference more than the linewidth of coherent phonon between probe and pump is used for violating the phase-matching for probe and seed; the other is called polarization-decoupling, used in this experimental demonstration. Here the frequencies of probe and pump are same for convenience, with a consequence of phase-matching between probe and seed. Hence, violating SAM conservation is a workable way to prevent their coupling, i.e., setting probe and seed in mutually orthogonal polarizations. As a result, the probe can only interact with the SBA-excited coherent phonons, and no Stokes beam will be generated via BAPA if pump or seed is shut down. Figure 1(b1, b2) show the timing of the beams, where pump is 1ns later than seed to arrive at Coupling Cell for suppressing interaction between pump and incoherent-phonon noise, and the probe is 1ns later than pump as well to ensure that the coherent phonon have been well generated by SBA. All beams' powers are below the threshold of SBS to avoid self-SBS noise, where the energy of pump, seed and probe are set to 2mJ, 0.5mJ and 2mJ, respectively, and the Stokes beam of 1.05mJ is obtained. More experimental details see Supplement 1.

In experiments, firstly, an OAM state of $\ell$=1, shown in Figs. 2(a1, a1i), is carried by seed and pump respectively for exciting OAM phonons via SBA. According to Eq. (3), the SBA-excited OAM phonon states should be $|0+1\rangle=|1\rangle$ and $|1+0\rangle=|1\rangle$, respectively, if they really exist. Then, to confirm these OAM phonons, a Gaussian probe beam is used to reconvert them from acoustic domain to optical domain via BAPA. According to Eq. (7), the output idler Stoke signals should be $|0+1-0\rangle=|1\rangle$ and $|1+0-0\rangle=|1\rangle$, respectively, and if using an OAM probe beam of $\ell$=1to extract the phonons generated by Gaussian pump and seed, the output Stokes signal should be $|0+0-1\rangle=|-1\rangle$. It can be easily seen that the experimental results shown in Figs. 2(a2/a2i-a4/a4i) are all in good agreement with the above analysis. Therefore, the reversible OAM transfer in photon-phonon conversion and corresponding OAM selection rules shown in Eq. (3) and (7) are verified. Particularly, the non-symmetrical uniformity of intensity patterns in Figs. 2(a2, a4), is due to a non-collinear parametric process in this experiment. More specifically, the optical axis of the output beam is slightly changed compared with the input beam, i.e., carrying off-axis vortex [33]. Secondly, a fractional OAM state of $\ell$=1.5 shown in Figs. 2(b1) is input from seed, pump and probe, where the SBA-excited OAM phonon states should be $\ell$=1.5, $\ell$=1.5 and $\ell$=0 respectively, according to Eq. (3). Note, the azimuthal index $\ell$ of Laguerre-Gauss mode can only be integer values, non-integer value in fact describes a superposition state, and here a ring gap facing bottom right arises from phase discontinuity associated with a string of alternating charge vortices [34, 35]. It can be seen that the ring gaps of corresponding output Stokes states shown in Figs. 2(b2-b4) face bottom right ($\ell$=1.5), bottom right ($\ell$=1.5) and left ($\ell$=-1.5), respectively, which means still complying with the OAM selection rule in Eq. (7). Thirdly, for demonstrating the OAM selection rule shown in Eq. (7) more intuitively, two beams of $\ell$=$p$=1 and $\ell$=$p$=0 are used as pump and seed, respectively, to excite an expected phonon state of $\ell$=$p$=1. Then, an OAM beam of $\ell$=1 is used as probe to generate Stokes beam, as shown in Figs. 2(c1-c4). The Stokes beam of $\ell$=0 and $p$=1 indicates that not only the azimuthal index $\ell$ but also the radial index $p$ are conserved in our experimental setup. It should be note, index $p$ is not always conserved in nonlinear process [19, 22], and more general case about this point will be further discussed in the future. The generation method of beams of $\ell$=1.5 and $\ell$=$p$=1 can be seen in Ref. [36, 37] and Experimental details in Supplement 1. Fourthly, a crucial point concerned in OAM-multiplexing system is that whether

OAM-multiplexing channels can be reversibly converted between photonic and phononic domains. Therefore, we implement a reversible conversion for OAM-multiplexed channels. Here, superposing OAM beams of $\ell=\pm2$ and $\ell=\pm3$ are used for simulating OAM-multiplexed channels respectively. It can be seen that the input seed states and output Stokes states for supposing beams, shown in Figs. 2(d1, d2) and 2(d3, d4), still keep the rule in Eq. (7). This result indicates that phononic OAM could be multi- and de-multiplexed as its photonic analogue, a crucial property for OAM-multiplexing system, and this is to be expected as phonons are quasi-Boson. Note, According to Eq. (3) and (7), if a probe is Gaussian beam, i.e., $\ell_b=0$, the phonon signals will be reconverted back to optical domain without any channel change, or else so-called channel switching will be performed. For example, a data-exchange can be achieved by using an OAM probe of $\ell=6$ to reconvert OAM-multiplexed phonon channels of $\ell=2$, 4 back and implementing once mirror reflection for output Stokes beam.

To prove the feasibility of using photon-phonon conversion based signal-processing scheme in OAM-multiplexing system, besides above demonstrations, another important question is that how long the OAM state could be maintained in phonon signals, and more important, what it will be like over time. So, finally, we further demonstrate the stability of OAM state in phononic domain, i.e., stability of OAM-channel labels. Here, an OAM state of $\ell=1$ is converted into phonon beam, and then be reconverted back to light beam after successively increasing delay times between SBA and BAPA. Figure 3 (a) shows the time sequences of the input beams, the presence of delay time (1-26ns) between pump and probe can insure that pump and seed have exited from coupling cell when probe entering, and here the maximum reconversion efficiency (output Stokes energy/input probe energy) and retrieved Stokes energy are about 21% and 0.424mJ, respectively, when the delay time is 1ns, as shown in Figs. 3 (b) and (c1). In contrast, in the case of no additional delay discussed above, as shown in Figs. 1(b1, b2), the photon-phonon coupling is enhanced due to SBA and BAPA simultaneous occurring in the cell, the reconversion efficiency is about 52%. An exponential decayed reconversion efficiency shown in Figs 3 (c1) indicates that the SBA-excited phonon signal begins to damp rapidly as soon as it is generated, as previously reported in Ref. [25, 26]. Beyond that, here the phonon signal will remain in its initial OAM state before completely disappearing, as shown in Fig. 3 (c2-c5). This implies that no model-crosstalk will be introduced when phonon signals propagate in a homogeneous and isotropic medium, just like optical OAM channels. In addition, although the phonon channels can only be sustained up to 26ns in this proof-of-principle experiment, the decay time of phonon signals can be remarkably extended to tens of μs by handling low-frequency long-life phonons in silicon photon-phonon circuit or micro-cavity via forward-SBS process [24-27]. Besides, a weak continuous probe light is able to sustain and enhance the signals in phonon domain via BAPA as required. However, it is worth noting that phonon is not a good candidate for storage but an intermediary for mediating interactions between photons, i.e., a data cache. Therefore, most of the time we hope its decay time as shorter as possible for an applied pursuit of high refresh rate. According to this requirement, data in phonon cache also could be quickly wiped by performing anti-Stoke type photon-phonon coupling, i.e., an up conversion between probe and phonons with generating anti-Stokes photons.

## 4. CONCLUSION

In conclusion, we have introduced the concept of OAM into photon-phonon coupling and used a simple experimental setup to demonstrate this proposal. A well-defined OAM state can be inter-converted between photonic and phononic domains via the controllable OAM transfer in SBA and BAPA. Our finding reveals the potential of using phonons to modulate data in photonic OAM channels. Owing to the reversible OAM photon-phonon conversion, the long-coherence-time and low-velocity phononic OAM signals can provide many important signal-processing functions and protocols required by OAM-multiplexing system, such as data cache, routing, switching, exchange and correction. Particularly note, the interconversion in this proof-of-principle study is demonstrated in a homogeneous and isotropic medium, which means we just take the first step of this proposal. At present, the optical OAM waveguide is still in an initial stage, for realizing the proposed signal-processing photon-phonon circuit in this paper, a waveguide only supporting OAM photons is obviously not enough. Nevertheless, a hybrid OAM photonic-phononic waveguide is worth expecting. More remarkably, the OAM selection rules shown in Eq. (3) and (7) largely depend on the fact that phonons carry no SAM, or in other words, SAM and OAM are independently conserved in photon-phonon conversion. This property, separating SAM and OAM while converting, can provide unique protocols for processing signals carried by cylindrical vector beams as well. And furthermore, to achieve the total angular momentum's state transfer between light and collective excitation, a reversible photon-magnon conversion in magnetic media is worthy to explore, where magnons can also be manipulated by a microwave signal and magnetic field, and during preparing this manuscript, some works about photon-magnon conversion have been reported on arXiv [38, 39].

## Acknowledgements

We are very grateful to Prof. Robert W. Boyd and Dr. Jeremy Upham from University of Ottawa, and Dr. Zhiyuan Zhou from University of Science and Technology of China, they reviewed our manuscript seriously and gave us many valuable suggestions. We also thank Dr. Xinmin Guo from Harbin Institute of Technology for illustration assistances

This work is supported by the National Natural Science Foundation of China (Grant No. 11574065, 61378003), the Key Programs of the Natural Science Foundation of Heilongjiang Province of China (Grant No. ZD201415).

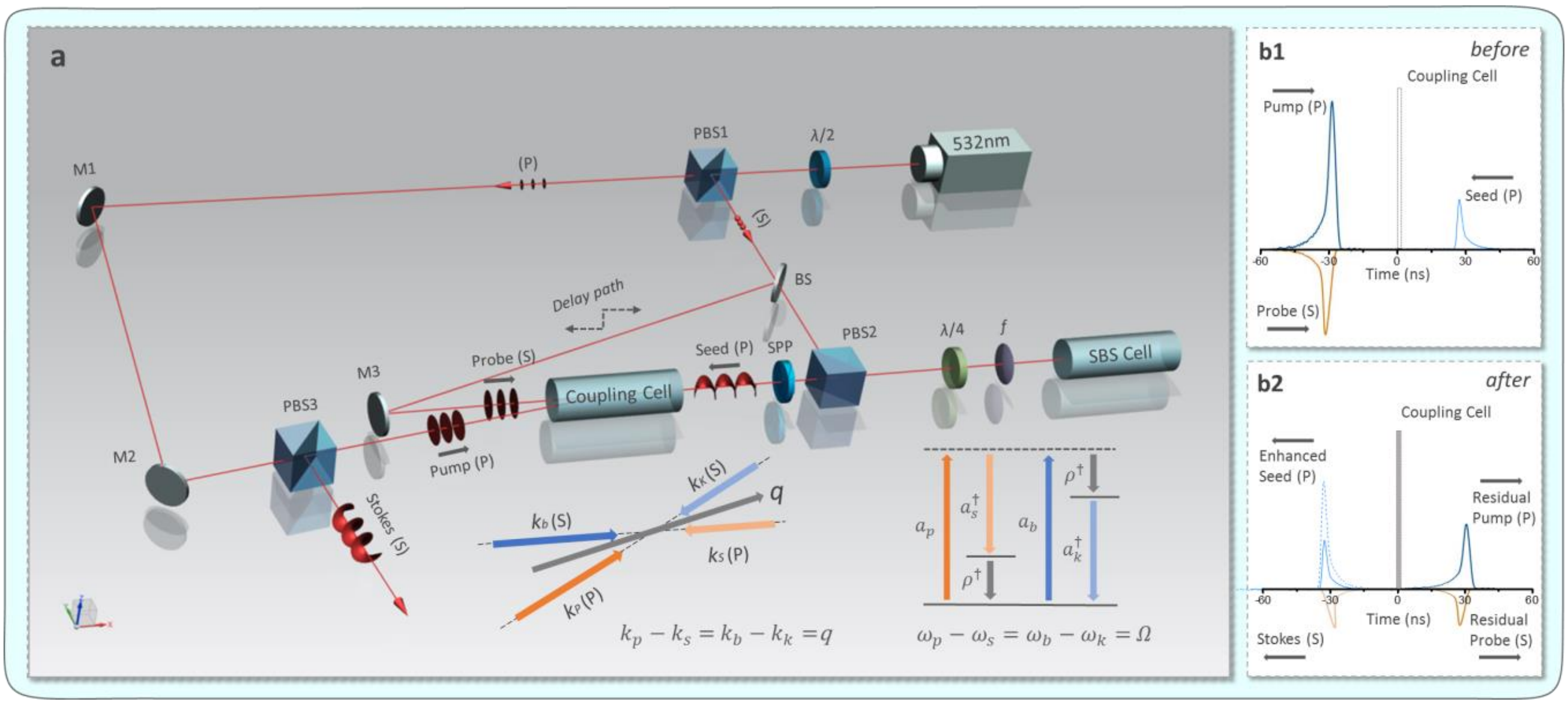

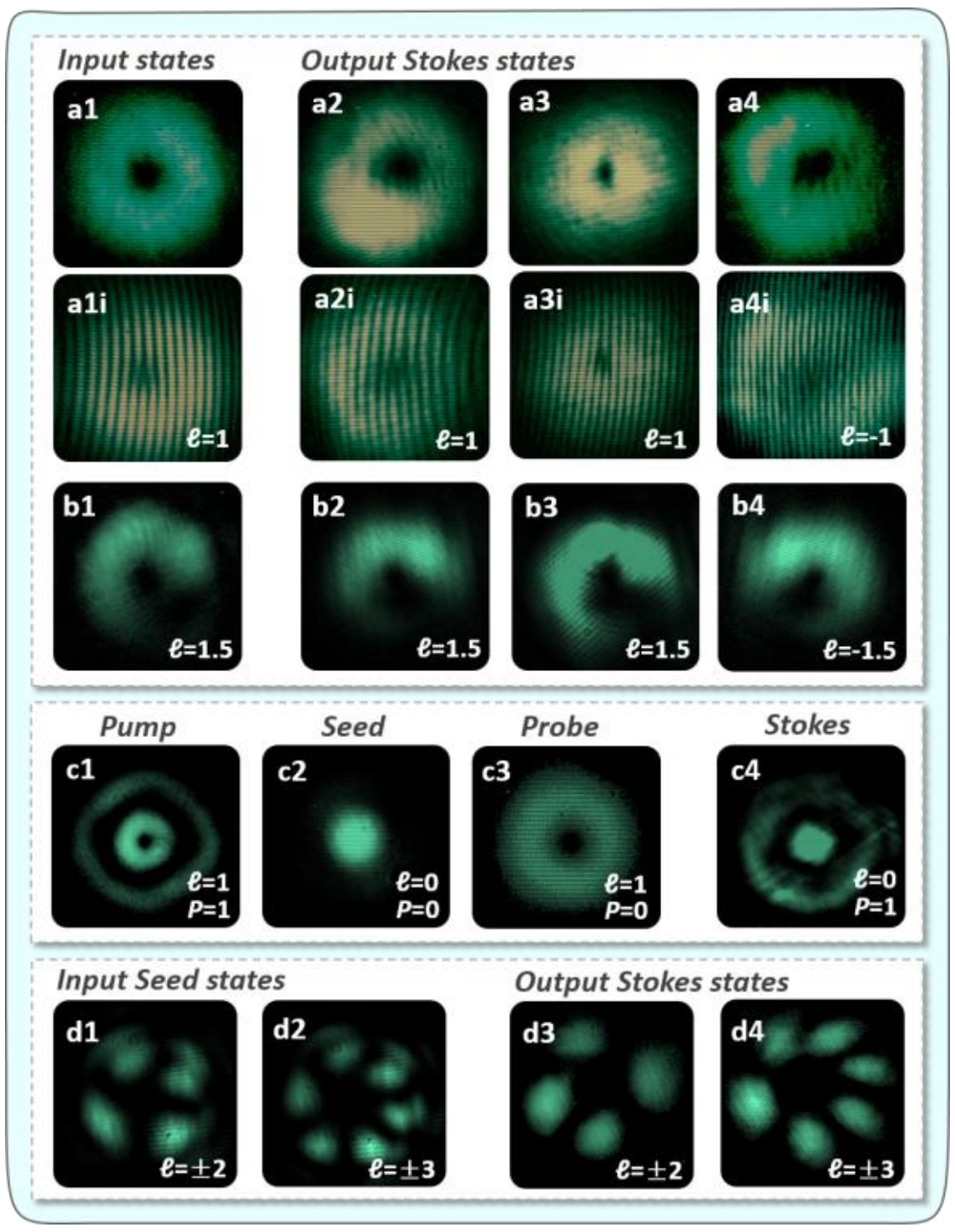

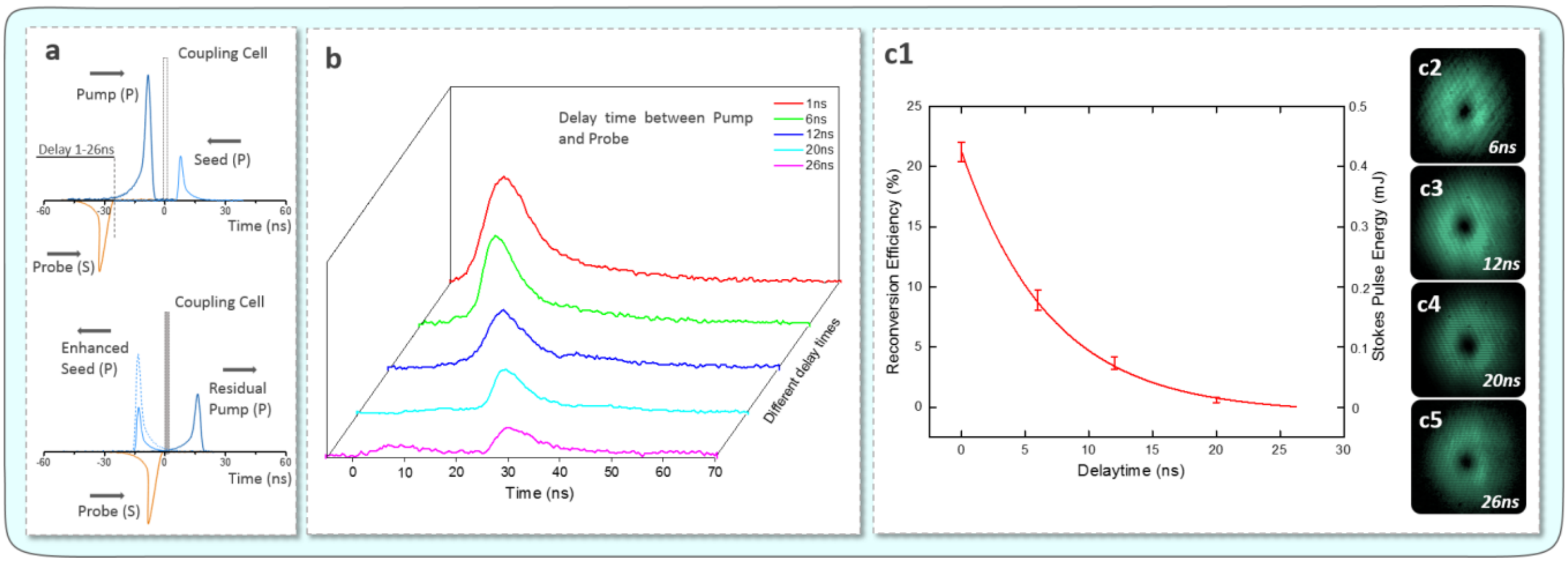

**Figure 1. Schematic presentation of reversible OAM photon-phonon conversion.** (a) Experimental setup. A λ/2 plate and a polarized beam splitter (PBS1) are used to vary the S and P contributions, and the P component is injected into the Coupling Cell as a Pump in SBA. The transmitted S component from BS is directed toward the SBS Cell to generate a P-polarized Stokes-frequency-shift seed beam, and the reflected S component from BS is used as a Probe in BAPA. The Pump and Seed interact quasi-collinearly in Coupling Cell via SBA to create a coherent phonon field, and then the Probe interacts with the phonon via BAPA to create a Stokes beam and reflect from PBS3. OAM beams are converted by spiral phase plates (SPP) or q-plate from Gaussian beams. Vector diagrams at right bottom describe the momentum and energy conservation relations of SBA and BAPA, respectively. (b1, b2) Timing of the beams before (b1) and after (b2) interactions in experiments

**Figure 2. Experimental results of reversible OAM photon-phonon conversion**. (a1-a4) Observed intensity profile/interferogram (a1/a1i) of input OAM beam of $\ell$=1 from seed, pump and probe respectively, and the intensity profiles/ interferograms (a2-a4/ a2i-a4i) of corresponding output Stokes beams. (b1-b4) Observed intensity profiles of input OAM beam of $\ell$=1.5 (b1) from seed, pump and probe respectively, and corresponding output Stokes beams (b2-b4). (c1-c4) Observed intensity profiles of input pump ($\ell$=$p$=1), seed ($\ell$=$p$=0), probe ($\ell$=1, $p$=0) and output Stokes ($\ell$=0, $p$=1) beams, respectively. (d1-d4) Observed intensity profiles of input seed beams $\ell$=±2/±3 (d1, d2) and corresponding output Stokes beams (d3, d4).

**Figure 3. Experimental results of OAM phonon signal's time stability**. (a) Timing of the delay experiment. Delay time between pump and probe is from 1ns to 26ns, the position of dotted line represents the case of 1ns delay. (b) Time-domain waveform of output Stokes beam versus delay time. (c1) Reconversion efficiency (output Stokes energy/input probe energy) and retrieved Stokes pulse energy versus delay time, the solid line is the exponential-decay fit. (c2-c5) Observed intensity profiles of output Stokes beams (phonon state is prepared in $\ell$=1) with different delays.